\documentclass[smallextended]{svjour3}     
\smartqed

\usepackage{graphicx}
\usepackage{amsmath,amssymb,harvard}
\newcommand{\bsnumber}{\begin{align}\begin{split}}
\newcommand{\X}{\mathcal{X}}

\journalname{Economic Theory}
\begin{document}

\title{Rearranging
Edgeworth-Cornish-Fisher Expansions \thanks{Chernozhukov and
Fern\'andez-Val gratefully acknowledge research support from the
National Science Foundation. Galichon's research is partly supported
by chaire X-Dauphine-EDF-Calyon ``Finance et D\'eveloppement
Durable''.}}



\author{Victor Chernozhukov  \and  Iv\'an Fern\'andez-Val  \and Alfred Galichon}


\institute{Victor Chernozhukov \at
          Massachusetts Institute of Technology, Department of Economics \& Operations Research
          Center, 50 Memorial Drive, Cambridge, 02142 Massachusetts,
          U.S.A. \\
          University College London, CEMMAP, London, U.K. \\
          \email{vchern@mit.edu}           
          \and
          Iv\'an Fern\'andez-Val \at
          Department of Economics, Boston University, 270 Bay State Road, Boston, 02215 Massachusetts, U.S.A. \\
          \email{ivanf@bu.edu}
          \and
          Alfred Galichon \at
          Department of Economics, Ecole Polytechnique, D$\acute{e}$partement d'Economie, 91128 Palaiseau Cedex, France\\
          \email{alfred.galichon@polytechnique.edu} }

\date{Received: date / Accepted: date}

\maketitle

\begin{abstract}
This paper applies a regularization procedure called increasing
rearrangement to monotonize Edgeworth and Cornish-Fisher expansions
and any other related approximations of distribution and quantile
functions of sample statistics. In addition to satisfying
monotonicity, required of distribution and quantile functions, the
procedure often delivers strikingly better approximations to the
distribution and quantile functions of the sample mean than the
original Edgeworth-Cornish-Fisher expansions.

\end{abstract}

\section{Introduction} \label{intro}

Approximations to the distribution of sample statistics of higher
order than the order $n^{-1/2}$ provided by the central limit
theorem are of central interest in the theory of asymptotic
statistics. See, e.g., \citeasnoun{bhattacharya_rao},
\citeasnoun{Rothenberg:handbook}, \citeasnoun{hall_bootstrap_book},
\citeasnoun{blinnikov_moessner}, \citeasnoun{vaart:text}, and
\citeasnoun{cramer_book}. An important tool for performing these
refinements is provided by the Edgeworth expansion
(\citeasnoun{Edgeworth_1905}, \citeasnoun{Edgeworth_1907}), which
approximates the distribution of the statistics of interest around
the limit distribution (often the normal distribution) using a
combination of Hermite polynomials with coefficients defined in
terms of population moments. Inverting the expansion yields a
related higher order approximation, the Cornish-Fisher expansion
(\citeasnoun{Cornish_Fisher_1938},
\citeasnoun{Fisher_Cornish_1960}), to the quantiles of the statistic
around the quantiles of the limiting distribution.

One important shortcoming of either the Edgeworth or Cornish-Fisher
expansions is that the resulting approximations to the distribution
and quantile functions are not necessarily increasing, which
violates an obvious monotonicity requirement. This comes from the
fact that the polynomials involved in the expansion are not
monotone. Here we propose to use a procedure, called the
rearrangement, to restore the monotonicity of the approximations
and, perhaps more importantly, to improve the estimation properties
of these approximations. The resulting improvement is due to the
fact that the rearrangement necessarily brings the non-monotone
approximations closer to the true monotone target function.

The main findings of the paper can be illustrated through a single
picture given in Figure 1, where we plot the true distribution
function of a standardized sample mean $X$ based on a small sample,
a third order Edgeworth approximation to that distribution, and the
rearrangement of the third order approximation.  We see that the
Edgeworth approximation is sharply non-monotone and provides a
rather poor approximation to the distribution function. The
rearrangement merely sorts the value of the approximate distribution
function in increasing order.  One can see that the rearranged
approximation, in addition to being monotonic, is a much better
approximation to the true function than the original approximation.

\begin{figure*}
\includegraphics[width=\textwidth,height=\textwidth]{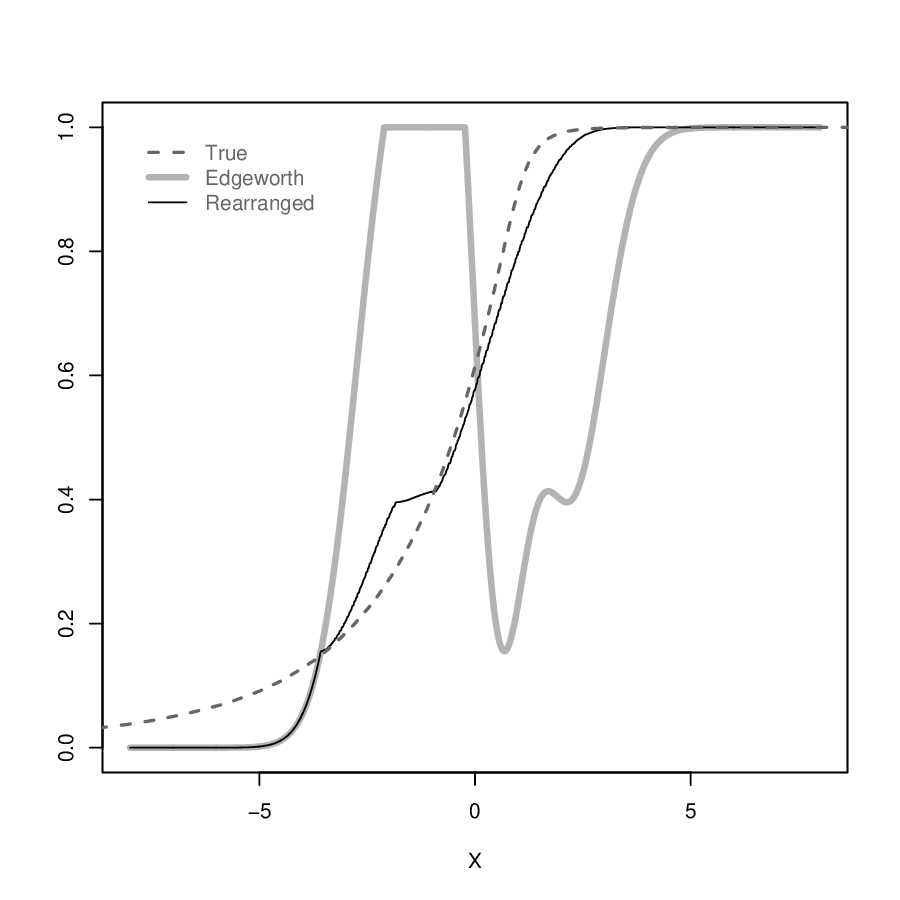}
\caption{Distribution function for the standardized sample mean of
Log-normal random variables (sample size 5), third order Edgeworth
approximation, and rearranged third order Edgeworth approximation.}
\label{Fig:1}       
\end{figure*}

We organize the rest of the paper as follows. In Section
\ref{sec:1}, we describe the rearrangement and qualify the
approximation property it provides for monotonic functions. In
Section \ref{sec:2}, we introduce the rearranged
Edgeworth-Cornish-Fisher expansions and explain how they produce
better approximations to distributions and quantiles of sample
statistics.  In Section \ref{sec:3}, we illustrate the procedure
with several additional examples.

\section{Improving Approximations of Monotone Functions
by Rearrangement} \label{sec:1}

In what follows, let $\mathcal{X}$ be a compact interval. We first
consider an interval of the form $\mathcal{X}=[0,1]$. Let $f(x)$ be
a measurable function mapping $\mathcal{X}$ to $K$, a bounded subset
of $\Bbb{R}$. Let $F_{f}(y) := \int_{\mathcal{X}} 1\{ f(u) \leq y \}
du$ denote the distribution function of $f(X)$ when $X$ follows the
uniform distribution on $[0,1]$. Let
$$f^*(x) = Q_f(x) := \inf \left \{ y \in \Bbb{R}:  F_{f}(y)
\geq x \right \}$$ be the quantile function of $F_f(y)$. Thus,
$$
f^*(x): = \inf \left \{ y \in \Bbb{R}: \left [ \int_{\mathcal{X}}
1\{ f(u) \leq y \} du \right] \geq x \right \}.
$$  This function $f^*$ is called the increasing
rearrangement of the function $f$.  The rearrangement is a tool
extensively used in functional analysis and optimal transportation
(see, e.g., \citeasnoun{HLP52} and \citeasnoun{villani}.) It
originates in the work of Chebyshev, who used it to prove a set of
inequalities (\citeasnoun{bronshtein_handbook_mathematics}, p. 31).
Here, we employ this tool to improve approximations of monotone
functions, such as the Edgeworth-Cornish-Fisher approximations to
the distribution and quantile functions of sample statistics.

The rearrangement operator simply transforms a function $f$ to its
quantile function $f^*$.  That is, $x\mapsto f^*(x)$ is the quantile
function of the random variable $f(X)$ when $X\sim U(0,1)$. Another
convenient way to think of the rearrangement is as a sorting
operation: Given values of the function $f(x)$ evaluated at $x$ in a
fine enough mesh of equidistant points, we simply sort the values in
increasing order. The function created in this way is the
rearrangement of $f$.

Finally, if $\mathcal{X}$ is of the form $[a,b]$ with $a < b$, let
$\bar x(x) = (x-a)/(b-a) \in [0,1]$ for $x \in \mathcal{X}$,
$x(\bar{x}) = a + (b-a)\bar{x} \in [a,b]$ for $\bar{x} \in [0,1]$,
and $\bar f^*$ be the rearrangement of the function $\bar f(\bar x)
= f(x(\bar x))$ defined on $\bar{ \mathcal{X}}=[0,1]$. Then, the
rearrangement of $f$ is defined as
$$
f^*(x) := \bar f^*(\bar x(x)).
$$


The following result establishes that the rearrangement always
improves the quality of the approximation to a monotone target
function.

\begin{proposition}[Improving Approximation of Monotone Functions]\label{betterapproximation}
Let $f_0: \mathcal{X}=[a,b] \to K$ be a weakly increasing measurable
function in $x$, where $K$ is a bounded subset of $\Bbb{R}$. This is
the target function that we want to approximate. Let $\widehat f:
\mathcal{X} \to K$ be another measurable function, an initial
approximation to the target function $f_0$.

\noindent 1. For any $p \in [1,\infty]$, the rearrangement of
$\widehat f$, denoted by $\widehat f^*$, weakly reduces the
estimation error:
 \begin{equation}\label{P11}
\left[\int_{\mathcal{X}} \left| \widehat f^*(x) - f_0(x) \right|^p d
x \right]^{1/p} \leq \left[ \int_{\mathcal{X}} \left| \widehat f(x)
- f_0(x) \right|^p d x \right]^{1/p}.
 \end{equation}

\noindent 2. Suppose that there exist regions $\X_0$ and $\X'_0$,
each of measure greater than $\delta>0$, such that for all $x \in
\X_0$ and $x' \in \X'_0$ we have that (i) $x'> x$, (ii) $\widehat
f(x) > \widehat f(x') + \epsilon$, and (iii) $f_0(x')
> f_0(x) + \epsilon$, for some $\epsilon >0$.  Then the
gain in the quality of approximation is strict for $p \in
(1,\infty)$. Namely, for any $p \in [1,\infty]$,
 \begin{equation}\label{P12}
\left[\int_{\mathcal{X}} \left| \widehat f^*(x) -f_0(x) \right|^p d
x \right]^{1/p} \leq \left[ \int_{\mathcal{X}} \left| \widehat f(x)
-f_0(x)\right|^p d x  - \delta_{\mathcal{X}} \eta_p \right]^{1/p},
\end{equation}
where $\eta_p = \inf\{ |v - t'|^p + |v' - t|^p - |v-t|^p - |v'-t'|^p
\}$ and $\eta_p>0$ for $p \in (1,\infty)$, with the infimum taken
over all $v, v', t, t'$ in the set $K$ such that $v' \geq v +
\epsilon$ and $t' \geq t + \epsilon$; and $\delta_{\mathcal{X}} =
\delta/(b-a)$.
\end{proposition}

\begin{corollary}[Strict Improvement]  \label{cor: strict_improvement}
If the target function $f_0$ is
increasing over $\mathcal{X}$ and $\widehat f$ is decreasing over a
subset of $\mathcal{X}$ that has positive measure, then the
improvement in $L_p$ norm, for $p \in (1, \infty)$, is necessarily
strict.
\end{corollary}

The first part of Proposition \ref{betterapproximation} states the
weak inequality (\ref{P11}), and the second part states the strict
inequality (\ref{P12}). As an implication, Corollary \ref{cor:
strict_improvement} states that the inequality is strict for $p \in
(1, \infty)$ if the initial approximation $\widehat f(x)$ is
decreasing on a subset of $\mathcal{X}$ having positive measure,
while the target function $f_0(x)$ is increasing on $\mathcal{X}$
(where by increasing, we mean strictly increasing throughout).
Proposition \ref{betterapproximation} establishes that the
rearranged approximation $\widehat f^*$ has a smaller estimation
error in the $L_p$ norm than the initial approximation whenever the
latter is not monotone. This is a very useful and generally
applicable property that is independent of the way the initial
approximation of $f_0$ is obtained.

\begin{remark} An indirect proof of the weak inequality
(\ref{P11}) is a simple but important consequence of the following
classical inequality due to \citeasnoun{lorentz:1953}: Let $q$ and
$g$ be two functions mapping $\mathcal{X}$ to $K$, a bounded subset
of $\Bbb{R}$. Let $q^{*}$ and $g^{*}$ denote their corresponding
increasing rearrangements. Then,
$$
\int_{\mathcal{X}} L(q^*(x), g^*(x)) d x  \leq \int_{\mathcal{X}}
L(q(x), g(x) ) dx,
$$
for any submodular discrepancy function $L: \Bbb{R}^2 \mapsto
\Bbb{R}$. Set $q(x) = \widehat f(x)$, $q^*(x) = \widehat f^*(x)$,
$g(x) = f_0(x)$, and $g^*(x) = f_0^*(x)$. Recall that a function $L:
\Bbb{R}^2 \mapsto \Bbb{R}$ is submodular if for each pair of vectors
$(v,t)$ and $(v',t')$ in $\Bbb{R}^2$, we have that $L(v\wedge v', t
\wedge t') + L(v \vee v', t \vee t') \leq L( v, t) + L(v', t').$
When the function $L$ is smooth, submodularity is equivalent to
$\partial^2 L(v,t)/ (\partial v
\partial t)  \leq 0$ holding for each $(v, t)$ in $\Bbb{R}^2$. Now, note that in our case
$ f_0^*(x) = f_0(x)$ almost everywhere, that is, the target function
is its own rearrangement. Moreover, $L(v,w ) = |w-v|^p$ is
submodular for $p \in [1, \infty)$.  This proves the first part of
the proposition above. For $p = \infty$, the first part follows by
taking the limit as $p \to \infty$.
\end{remark}

\begin{remark} The following immediate implication of the above
finite-sample result is also worth emphasizing: The rearranged
approximation $\widehat f^*$ inherits the $L_p$ rates of convergence
from the initial approximations $\widehat f$. For $p \in
[1,\infty]$, if $\lambda_n = [ \int_{\mathcal{X}} | \widehat f(x) -
f_0(x) |^p d u ]^{1/p} = O_P(a_n)$ for some sequence of constants
$a_n$, then $ [\int_{\mathcal{X}} | \widehat f^*(x) - f_0(x)  |^p d
x ]^{1/p} \leq \lambda_n = O_P(a_n)$. However, while the rate is the
same, the error itself is smaller.
\end{remark}

\begin{remark}
One of the following methods can be used for computing the
rearrangement. Let $\{X_j, j=1,...,B\}$ be either (1) a set of
equidistant points in $[0,1]$ or (2) a sample of i.i.d. draws from
the uniform distribution on $[0,1]$. Then the rearranged
approximation$\hat f^*(u)$ at point $u \in \mathcal {X}$ can be
approximately computed as the $u$-quantile of the sample $ \{
f(X_j), j=1,...,B\}$. The first method is deterministic, and the
second is stochastic. Thus, for a given number of draws $B$, the
complexity of computing the rearranged approximation $f^*(u)$ in
this way is equivalent to the complexity of computing the sample
$u$-quantile in a sample of size $B$. The number of evaluations $B$
can depend on the problem. Suppose that the density function of the
random variable $f(X)$, when $X \sim U(0,1)$, is bounded away from
zero over a neighborhood of $f^*(x)$. Then $f^*(x)$ can be computed
with the accuracy of $O_P(1/\sqrt{B})$, as $B \to \infty$, where the
rate follows from the results of \citeasnoun{knight:limits}.
\end{remark}

\begin{remark} One can also consider \textit{weighted
rearrangements} that give different importance to different areas of
the curve. Indeed, consider an absolutely continuous distribution
function $\Lambda$ on $\mathcal{X}=[a,b]$, then  we have that  for $
(\widehat f \circ \Lambda^{-1}) ^* (u)$ denoting the rearrangement
of $u\mapsto \widehat  f ( \Lambda^{-1}(u))$
\begin{multline*}
\left[\int_{[0,1]} \left|  ( \widehat f \circ \Lambda^{-1}) ^* (u) -
f_0 (\Lambda^{-1} (u)) \right|^p d u \right]^{1/p} \\
 \leq \left[
\int_{[0,1]}  \left| \widehat f (\Lambda^{-1} (u)) - f_0
(\Lambda^{-1} (u)) \right|^p d u \right]^{1/p},
\end{multline*}
or, equivalently, by a change of variable, setting
$$
\widehat f^*_{\Lambda}(x)  =  (\widehat f \circ \Lambda^{-1}) ^*
(\Lambda (x))
$$
we have that
 \begin{equation*}
\left[\int_{\mathcal{X}} \left| \widehat f^*_{\Lambda}(x) -
 f_0(x) \right|^p d \Lambda(x) \right]^{1/p}
\leq \left[ \int_{\mathcal{X}} \left| \widehat f(x) - f_0(x)
\right|^p  d \Lambda(x) \right]^{1/p}.
 \end{equation*}
Thus, the function $x \mapsto \widehat f^*_{\Lambda}(x)$ is the
weighted rearrangement that improves over the initial approximation
in the norm that is weighted according to the distribution function
$\Lambda$. At the same time it is important to note that the
(weighted) rearrangement does not necessarily guarantee improvements
at every point or any specific point.
\end{remark}

\noindent   \textbf{Proof of Proposition \ref{betterapproximation}.}
We consider the case where $\mathcal{X} =[0,1]$ only, as the more
general intervals can be dealt similarly. The first part establishes
the weak inequality, following, in part, the strategy in Lorentz's
proof. The proof focuses directly on obtaining the result stated in
the proposition. The second part establishes the strong inequality.

Proof of Part 1.  We assume first  that the functions $\hat f$ and
$f_0$ are step functions, constant on intervals $((s-1)/r, s/r]$,
$s=1,\ldots,r$. For each step function $f$ with $r$ steps we
associate an $r$-vector $f$ whose $s$-th element, denoted $f_{s}$,
equals to the value of function $f$ on the $s$-th interval, and vice
versa. Let us define the sorting operator $S$ acting on vectors (and
functions) $f$ as follows. Let $k$ be an integer in $1,\ldots,r$
such that $f_k > f_m$ for some $m>k$. If $k$ does not exist, set $Sf
= f$. If $k$ exists, set $Sf$ to be a $r$-vector with the $k$-th
element equal to $f_m$, the $m$-th element equal to $f_k$, and all
other elements equal to the corresponding elements of $f$. Finally,
given a vector $Sf$ there is a step function $Sf$ associated to it,
as stated above.

For any submodular function $L: \Bbb{R}^2 \to \Bbb{R}_+$, by $f_k
\geq f_m$, $f_{0m} \geq f_{0k}$ and the definition of the
submodularity, $L(f_m, f_{0k}) + L (f_k, f_{0m}) \leq $ $L(f_k,
f_{0k}) + L(f_m, f_{0m}).$  Thus conclude that $ \int_{\X} L\{S\hat
f(x), f_0(x)\} dx \leq \int_{\X} L\{ \hat f(x), f_0(x) \}dx,$ using
that  we integrate step functions. Applying the sorting operator a
sufficient finite number of times to $\hat f$, we obtain a
completely sorted, that is, rearranged, vector $\hat f^*$. Thus, we
can express $\hat f^*$ as  $\hat f^*= S \ldots S \hat f$, where the
operator $S$ is applied finitely many times. By repeating the
argument above, each application weakly reduces the estimation
error. Therefore,
\begin{equation}\label{maininequality}
\int_\X L\{\hat f^*(x), f_0(x)\} d x \leq \int_{\X} L\{ S \ldots
S\hat f(x) , f_0(x) \} dx \leq \int_{\X} L\{\hat f(x), f_0(x)\} dx.
 \end{equation}

Next we extend this result to general measurable functions $\hat f$
and $f_0$ mapping $[0,1]$ to $K$, where $f_0$ is a quantile
function. Take a subsequence of  bounded step functions $\hat
f^{(q)}$ and $f_0^{(q)}$, with $f_0^{(q)}$ being quantile functions,
converging to $\hat f$ and $f_0$ almost everywhere as index $q \to
\infty$ along an increasing sequence of integers. The almost
everywhere convergence of $\hat f^{(q)}$ to $\hat f$ implies the
almost everywhere convergence of its quantile function $\hat
f^{*(q)}$ to the quantile function of the limit, $\hat f^*$
(\citeasnoun{vaart:text}, p. 305). Since (\ref{maininequality})
holds for each $q$ along the sequence, the dominated convergence
theorem implies that (\ref{maininequality}) also holds for the
general case.

It remains to show the existence of the subsequence in the preceding
paragraph. Using series expansion in the Haar basis, any function in
$L^2[0,1]$ can be approximated in $L^2$ norm  by a sequence of
$r$-step functions, where $r=2^j-1$ and $j =1,\ldots,\infty$
\cite{pollard:measure}. Hence there is a sequence  of step functions
$ \hat f^{(r)}$ and $f^{(r)}_0$ converging to $\hat f$ and $f_0$ in
$L^2$ norm; the functions in the sequence necessarily take values in
$K$; by \citeasnoun{pollard:measure}, p. 38, we can extract a
further subsequence $\hat f^{(q)}$ and $f^{(q)}_0$, with $q$ running
over an increasing sequence of integers, converging to $\hat f$ and
$f_0$ almost everywhere. Finally, replace $f^{(q)}_0$ by their
quantile functions, i.e., rearrangements, which retain the almost
everywhere convergence property to $f_0$ by \citeasnoun{vaart:text},
p. 305. \qed

Proof of Part 2. Consider the step functions, as defined in the
proof of Part 1.  By setting $r$ sufficiently large, we  can take
them to satisfy the following hypotheses: there exist regions $\X_0$
and $ \X'_0$, each of measure greater than $\delta>0$, such that for
all $x \in \X_0$ and $x' \in \X_0'$, we have that (i) $x'
> x$, (ii) $\hat f(x) > \hat f(x') + \epsilon$, and (iii)
$f_0(x') > f_0(x) + \epsilon$, for $\epsilon>0$ specified in the
proposition.   For any strictly submodular function $L: \Bbb{R}^2
\to \Bbb{R}_+$ we have that $\eta = \inf \{ L(v',t) + L(v,t') -
L(v,t) - L(v',t') \} >0, $ where the infimum is taken over all $v,
v', t, t'$ in the set $K$ such that $v' \geq v + \epsilon$ and $t'
\geq t + \epsilon$.  We can begin sorting by exchanging an element
$\hat f(x)$, $x \in \X_0$, of $r$-vector $\hat f$ with an element
$\hat f(x')$, $x' \in \X_0'$, of $r$-vector
 $\hat f$. This induces a sorting gain of at
least $\eta$ times $1/r$.  The total mass of points that can be
sorted in this way is at least $\delta$. We then proceed to sort all
of these points in this way, and then continue with the sorting of
other points. After the sorting is completed, the total gain from
sorting is at least $\delta \eta$. That is, $ \int_{\X} L\{\hat
f^*(x), f_0(x)\}dx  \leq \int_{\X} L\{\hat f(x), f_0(x)\} dx -
\delta \eta.$

We then extend this inequality to the general measurable functions
exactly as in the proof of Part 1. \qed

In the next section, we apply rearrangements to improve the
Edgeworth-Cornish-Fisher and related approximations to distribution
and quantile functions.

\section{Improving Edgeworth-Cornish-Fisher and Related expansions} \label{sec:2}

\subsection{Improving Quantile Approximations by Rearrangement} \label{subsec:2.1}

We first consider the quantile case. Let $Q_n$ be the quantile
function of a statistic $X_n$, i.e.,
$$
Q_n(u) = \inf\{ x \in \Bbb{R}: Pr[X_n \leq x] \geq u \},
$$
which we assume to be strictly increasing. Let $\widehat Q_n$ be an
approximation to the quantile function $Q_n$ satisfying the
following relation:
 \bsnumber \label{CF1}
Q_n(u) = \widehat Q_n(u) + \epsilon_n(u), \ \ |\epsilon_n(u) | \leq
a_n, \ \ \text{ for all } u \in \mathcal{U}_n,
 \end{split}\end{align}
where $a_n$ is some sequence of positive numbers going to zero as $n
\rightarrow \infty$, $\mathcal{U}_n=[\varepsilon_n, 1-
\varepsilon_n] \subseteq [0,1]$, and $\varepsilon_n< 1$ is some
sequence of positive numbers possibly going to zero as $n
\rightarrow \infty$. For example, (\ref{CF1}) holds with
$\varepsilon_n = n^{-c}$, $c>0$, under certain conditions on the
moments (e.g., \citeasnoun{hall_bootstrap_book}).

The leading example of such an approximation is the inverse
Edgeworth, or Cornish-Fisher, expansion of the quantile function of
a sample mean.  If $X_n$ is the standardized sample mean, $X_n =
n^{-1/2} \sum_{i=1}^n (Y_i - E[Y_i])/\sqrt{Var(Y_i)}$, based on a
random sample $(Y_1,..., Y_n)$ of $Y$, then we have the following
$J$-th order expansion
 \bsnumber \label{CF-mean} & Q_n(u)  = \widehat Q_n(u) + \epsilon_n(u), \\
& \widehat Q_n(u)  =
R_1(\Phi^{-1}(u)) + R_2(\Phi^{-1}(u))/n^{1/2} + ... + R_J(\Phi^{-1}(u))/n^{(J-1)/2}, \\
\ \  & | \epsilon_n(u) |  \leq C n^{-J/2}, \ \ \text{ for all } u
\in \mathcal{U}_n = [\varepsilon_n, 1- \varepsilon_n],
\\ & \text{ for some $\varepsilon_n \searrow 0$  and $C
>0$},
 \end{split}\end{align}
provided that a set of regularity conditions, specified, e.g., in
\citeasnoun{zolotarev_pobabilistic}, hold. Here $\Phi$ and
$\Phi^{-1}$ denote the distribution function and quantile function
of a standard normal random variable. The first three terms of the
expansion are given by the polynomials,
\bsnumber\label{CF-polynomials}
& R_{1}(z) = z, \\
& R_{2}(z) = \lambda (z^2 - 1)/6, \\
& R_{3}(z) = (3 \kappa (z^3 - 3z) - 2\lambda^2 (2 z^3 - 5z))/72,
\end{split}\end{align}
where $\lambda$ is the skewness and $\kappa$ is the kurtosis of the
random variable $Y$. The Cornish-Fisher expansion is one of the
central approximations of the asymptotic statistics. Unfortunately,
an inspection of the expressions for the polynomials in
(\ref{CF-polynomials}) reveals that this expansion does not
generally deliver a monotone approximation of the quantile function.
This shortcoming has been pointed and discussed in detail for
example by \citeasnoun{hall_bootstrap_book}.   The nature of the
polynomials is such that there always exists a large enough range
$\mathcal{U}_n$ over which the Cornish-Fisher approximation is not
monotone, cf., \citeasnoun{hall_bootstrap_book}. As an example, in
the case of the second order approximation ($J=2$) we have that, for
$\lambda <0$
\begin{equation*}
\widehat Q_n (u) \searrow - \infty,  \text{ as } u \nearrow 1,
 \end{equation*}
that is, the Cornish-Fisher ``quantile" function $\widehat Q_n$ is
decreasing far enough in the tails. This example merely suggests a
potential problem that may or may not apply to practically relevant
ranges of probability indices $u$. Indeed, specific numerical
examples given below show that in small samples the non-monotonicity
can occur in practically relevant ranges. Of course, in sufficiently
large samples, the regions of non-monotonicity are squeezed quite
far into the tails.

 Let $\widehat Q_n^*$ be the rearrangement of $\widehat Q_n$. Then we have that for any $p \in [1,\infty]$, the rearranged quantile function reduces the approximation error of the
original approximation:
 \begin{eqnarray}\label{CF-improve}
\left[\int_{\mathcal{U}_n} \left| \widehat Q^*_n(u) - Q_n(u)
\right|^p d u \right]^{1/p} &\leq& \left[ \int_{\mathcal{U}_n}
\left|
\widehat Q_n(u) - Q_n(u) \right|^p d u \right]^{1/p} \notag \\
&\leq& (1-2\varepsilon_n)^{1/p} \ a_n,
 \end{eqnarray}
with the first inequality holding strictly for $p \in (1,\infty)$
whenever $\widehat Q_n$ is decreasing on a region of $\mathcal{U}_n$
of positive measure.  We can give the following probabilistic
interpretation to this result. Under condition (\ref{CF1}), there
exists a variable $U= F_n(X_n)$, where $F_{n}$ is the distribution
function of $X_{n}$, such that both the stochastic expansion
 \bsnumber\label{CF-probabilistic1}
X_n  = \widehat Q_n(U) + O_P(a_n),
 \end{split}\end{align}
and the expansion \bsnumber\label{CF-probabilistic2}
 X_n  = \widehat Q_n^*(U) +
O_P(a_n),
 \end{split}\end{align}
hold.\footnote{$\widehat Q_n^*(U)$ is defined only on
$\mathcal{U}_n$, so we can set $\widehat Q_n^*(U)=Q_n(U)$ outside
$\mathcal{U}_n$, if needed. Of course, $U \not \in \mathcal{U}_n$
with probability going to zero if $\varepsilon_{n} \searrow 0$ as $n
\rightarrow \infty$.} However, the variable $\widehat Q_n^*(U)$ in
(\ref{CF-probabilistic2}) is a better coupling to the statistic
$X_n$ than $\widehat Q_n(U)$ in (\ref{CF-probabilistic1}), in the
following sense: For each $p \in [1, \infty]$,
 \bsnumber\label{CF-better probabilistic}
\left\{E [1_n \cdot | \widehat Q_n^*(U) - X_n |^{p} \ ]
\right\}^{1/p} \leq \left\{ E [ 1_n \cdot | \widehat Q_n(U) - X_n
|^{p} \ ] \right\} ^{1/p},
 \end{split}\end{align}
where $1_n =1\{ U \in \mathcal{U}_n\}$. Indeed, property
(\ref{CF-better probabilistic}) immediately follows from
(\ref{CF-improve}).

The above improvements apply in the context of the sample mean
$X_n$. In this case, the probabilistic interpretation above is
directly connected to the higher order central limit theorem of
\citeasnoun{zolotarev_pobabilistic}, which states that under
(\ref{CF-mean})  we have the following higher-order probabilistic
central limit theorem,
\begin{equation*}
X_n = \widehat Q_n(U) + O_P(n^{-J/2}).
\end{equation*}
The term $\widehat Q_n(U)$ is Zolotarev's higher-order refinement
over the first order normal term $\Phi^{-1}(U)$.
\citeasnoun{sun_loader_mccormick} employ an analogous higher-order
probabilistic central limit theorem to improve the construction of
confidence intervals.

The application of the rearrangement to Zolotarev's term actually
delivers a clear improvement in the sense that it also leads to a
probabilistic higher order central limit theorem
\begin{equation*}
X_n = \widehat Q^*_n(U) + O_P(n^{-J/2}),
\end{equation*}
where the leading term $\widehat Q^*_n(U)$ is closer to $X_n$ than
Zolotarev's term $Q_n(U)$, in the sense of (\ref{CF-better
probabilistic}).

We summarize the above discussion into a formal proposition.

\begin{proposition} \label{prop: quantile_improvement} If the
expansion (\ref{CF1}) holds, then the improvement
 (\ref{CF-improve}) necessarily holds. The improvement
is necessarily strict if $\widehat Q_n$ is decreasing over a region
of $\mathcal{U}_n$ that has a positive measure. In particular, this
improvement property applies to the inverse Edgeworth approximation
to the quantile function of the sample mean, defined in
(\ref{CF-mean}).
\end{proposition}

\subsection{Improving Distributional Approximations by
Rearrangement} \label{subsec:2.2}

We next consider distribution functions. Let $F_n(x)$ be the
distribution function of a statistic $X_n$, assumed to be strictly
increasing, and $\widehat F_n(x)$ be an approximation to this
distribution such that the following relation holds:
\begin{equation}\label{E1}
F_n(x) = \widehat F_n(x) + \epsilon_n(x), \ \ | \epsilon_n(x) | \leq
a_n, \text{ for all } x \in \mathcal{X}_n,
 \end{equation}
where $a_n$ is some sequence of positive numbers going to zero as $n
\rightarrow \infty$, and $\mathcal{X}_n =[- b_n, c_n]$ is an
interval in $\Bbb{R}$ for some sequences of positive scalars $b_n$
and $c_n$ possibly growing to infinity. The choice of $b_n$ and
$c_n$ is unrestricted under certain conditions on the moments (e.g.,
 \citeasnoun{hall_bootstrap_book}).

The leading example of such an approximation is the Edgeworth
expansion of the distribution function of the sample mean. If $X_n$
is the standardized  sample mean, $X_n = n^{-1/2} \sum_{i=1}^n (Y_i
- E[Y_i])/\sqrt{Var(Y_i)}$, based on a random sample $(Y_1,...,
Y_n)$ of $Y$, then we have the following $J$-th order expansion
 \bsnumber \label{E-mean} & F_n(x)  = \widehat F_n(x) + \epsilon_n(x), \\
& \widehat F_n(x)  =
P_1(x) + P_2(x)/n^{1/2} + ... + P_J(x)/n^{(J-1)/2}, \\
\ \  & | \epsilon_n(x) |  \leq C n^{-J/2}, \ \ \text{ for all } x
\in \mathcal{X}_n,
 \end{split}\end{align}
for some $C
>0$, provided that a set of regularity conditions,
specified, e.g., in \citeasnoun{hall_bootstrap_book}, hold.  The
first three terms of the approximation are given by
\begin{align*}\begin{split}
& P_{1}(x) = \Phi(x), \\
& P_{2}(x) = - \lambda (x^2 - 1) \phi(x) / 6, \\
& P_{3}(x) = - (3 \kappa (x^3 - 3x) + \lambda^2 (x^5 - 10x^3 + 15x))
\phi(x) / 72,
\end{split}\end{align*}
where $\Phi$ and $\phi$ denote the distribution function and density
function of a standard normal random variable, and $\lambda$
 and $\kappa$ are the skewness and kurtosis of the random variable $Y$,
 respectively. Here too, the Edgeworth expansion is one of the central
approximations of the asymptotic statistics. Unfortunately, like the
Cornish-Fisher expansion, it generally does not provide a monotone
approximation of the distribution function. This shortcoming has
been pointed and discussed in detail by \citeasnoun{barton_dennis},
\citeasnoun{draper_tierney}, \citeasnoun{sargan_edgeworth}, and
\citeasnoun{balitskaya_zolotukhina}, among others.

 Let $\widehat F_n^*$ be the rearrangement of $\widehat F_n$.
 Then, we have that for any $p \in [1,\infty]$, the rearranged Edgeworth
 approximation reduces the approximation error of the
original Edgeworth approximation:
 \begin{eqnarray}\label{E-improve}
\left[\int_{\mathcal{X}_n} \left| \widehat F^*_n(x) - F_n(x)
\right|^p d x \right]^{1/p} &\leq& \left[ \int_{\mathcal{X}_n}
\left| \widehat F_n(x) - F_n(x) \right|^p d x \right]^{1/p} \notag
\\ &\leq& (c_n + b_n)^{1/p} \ a_n,
 \end{eqnarray}
with the first inequality holding strictly for $p \in (1,\infty)$
whenever $\widehat F_n$ is decreasing on a region of $\mathcal{X}_n$
of positive measure.

\begin{proposition} \label{prop: cdf_improvement} If expansion (\ref{E1}) holds, then the improvement
 (\ref{E-improve}) necessarily holds. The improvement
is necessarily strict if $\widehat F_n$ is decreasing over a region
of $\mathcal{X}_n$ that has a positive measure. In particular, this
improvement property applies to the Edgeworth approximation to the
distribution function of the sample mean, defined in (\ref{E-mean}).

\end{proposition}

\subsection{Weighted Rearrangement of Cornish-Fisher and
Edgeworth Expansions} \label{subsec:2.3}

In some cases, it can be worthwhile to weigh different areas of the
support differently than the Lebesgue (flat) weighting prescribes.
For example, it may be desirable to rearrange $\widehat F_n$ using
$F_n$ as a weighting measure. Indeed, by using $F_n$ as a weight, we
obtain a better matching with the P-value: $ P=F_n(X_n)$ (in this
quantity, $X_n$ is drawn according to the true $F_n$). Using such a
weight will provide a probabilistic interpretation for the
rearranged Edgeworth expansion, analogous to the probabilistic
interpretation for the rearranged Cornish-Fisher expansion. Although
the weight (true $F_n$) is not available, we can use the standard
normal measure $\Phi$ as the weighting measure instead. We may also
construct an initial rearrangement with the Lebesgue weight, and use
it as weight itself for a further weighted rearrangement (and even
continue to iterate in this fashion). Using non-Lebesgue weights may
also be desirable when we want the improved approximations to weigh
the tails more heavily. Whatever the reason may be for further
non-Lebesgue weighting, they have the following properties, which
follow immediately in light of Remark 3.

Let $\Lambda$ be a distribution function that admits a positive
density with respect to the Lebesgue measure on the region
$\mathcal{U}_n=[\varepsilon_n, 1- \varepsilon_n]$ for the quantile
case and on the region $\mathcal{X}_n=[-b_n, c_n]$ for the
distribution case. Then, if (\ref{CF1}) holds, the
$\Lambda$-weighted rearrangement $\widehat Q^*_{n, \Lambda}$ of the
function $\widehat Q_n$ satisfies
 \begin{eqnarray*}\label{CF-improve-weighted}
\left[\int_{\mathcal{U}_n} \left| \widehat Q^*_{n, \Lambda} (u) -
Q_n(u) \right|^p d \Lambda(u) \right]^{1/p} & \leq &  \left[
\int_{\mathcal{U}_n} \left| \widehat Q_n(u) - Q_n(u) \right|^p d
\Lambda(u)
\right]^{1/p} \\
&\leq & (\Lambda[1-\varepsilon_n]- \Lambda[\varepsilon_n])^{1/p} \
a_n,
 \end{eqnarray*}
where the first equality holds strictly when $\widehat Q$ is
decreasing on a subset of positive $\Lambda$-measure.  Furthermore,
if (\ref{E1}) holds, then the $\Lambda$-weighted rearrangement
$\widehat F^*_{n, \Lambda}$ of the function $\widehat F_n$ satisfies
 \begin{eqnarray*}\label{E-improve-weighted}
\left[\int_{\mathcal{X}_n} \left| \widehat F^*_{n,\Lambda} (x) -
F_n(x) \right|^p d \Lambda(x) \right]^{1/p} & \leq &  \left[
\int_{\mathcal{X}_n} \left| \widehat F_n(x) - F_n(x) \right|^p d
\Lambda(x)
\right]^{1/p} \\
&  \leq &  (\Lambda[c_n]- \Lambda[- b_n])^{1/p} \ a_n.
 \end{eqnarray*}

\section{Numerical Examples} \label{sec:3}
In addition to the Log-normal example given in the introduction, we
use the Gamma distribution to illustrate the improvements that the
rearrangement provides.  Let $(Y_1,...,Y_n)$ be an i.i.d. sequence
of Gamma(1/16,16) random variables. The statistic of interest is the
standardized sample mean $X_n = n^{-1/2} \sum_{i=1}^n (Y_i -
E[Y_i])/\sqrt{Var(Y_i)}$. We consider samples of sizes $n=4, 8, 16$,
and $32$. In this example, the distribution function $F_n$ and
quantile function $Q_n$ of the statistic $X_n$ are available in a
closed form, making it easy to compare them to the Edgeworth
approximation $\widehat F_n$ and the Cornish-Fisher approximation
$\widehat Q_n$, as well as to the rearranged Edgeworth approximation
$\widehat F_n^*$ and the the rearranged Cornish-Fisher approximation
$\widehat Q_n^*$. For the Edgeworth and Cornish-Fisher
approximations, as defined in the previous section, we consider
third order expansions, that is we set $J=3$.


\begin{figure*}[h]
\includegraphics[width=\textwidth,height=1.25\textwidth]{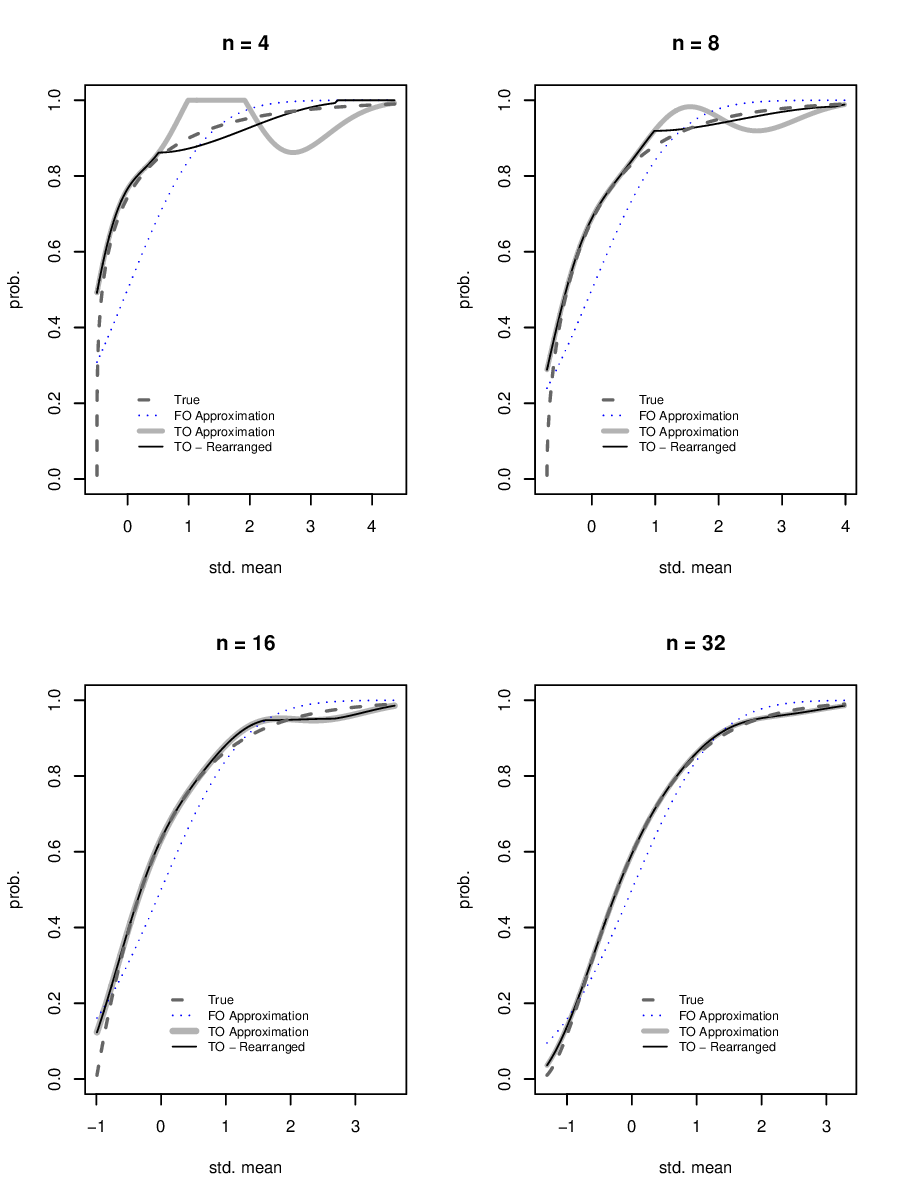}
\caption{Distribution Functions, First Order Approximations, Third
Order Approximations, and Rearrangements for the standardized sample
mean from a Gamma(1/16, 16) population.}
\label{fig:2}       
\end{figure*}


%

\begin{table}[h]
\caption{Estimation errors for approximations to the distribution
function of the standardized sample mean from a Gamma(1/16, 16)
population.\label{table1}}
\begin{tabular}{lllll} \hline\noalign{\smallskip}
\multicolumn{1}{l}{
}&\multicolumn{1}{c}{First Order }&\multicolumn{1}{c}{Third Order }&\multicolumn{1}{c}{Rearranged - TO}&\multicolumn{1}{c}{Ratio (RTO/TO)}\\
\noalign{\smallskip}\hline\noalign{\smallskip}
\multicolumn{1}{l}{ }&\multicolumn{4}{c}{$n = 4$}  \\
$L_1$&0.07&0.05&0.02&0.38\\
$L_2$&0.10&0.06&0.03&0.45\\
$L_3$&0.13&0.07&0.04&0.62\\
$L_4$&0.15&0.08&0.07&0.81\\
$L_\infty$&0.30&0.48&0.48&1.00\\
\multicolumn{1}{l}{ }&\multicolumn{4}{c}{$n = 8$}  \\
$L_1$&0.06&0.03&0.01&0.45\\
$L_2$&0.08&0.04&0.02&0.63\\
$L_3$&0.10&0.05&0.04&0.85\\
$L_4$&0.11&0.06&0.06&0.96\\
$L_\infty$&0.23&0.28&0.28&1.00\\
\multicolumn{1}{l}{ }&\multicolumn{4}{c}{$n = 16$}  \\
$L_1$&0.05&0.01&0.01&0.97\\
$L_2$&0.06&0.02&0.02&0.99\\
$L_3$&0.08&0.03&0.03&1.00\\
$L_4$&0.08&0.04&0.04&1.00\\
$L_\infty$&0.15&0.11&0.11&1.00\\
\multicolumn{1}{l}{ }&\multicolumn{4}{c}{$n = 32$}  \\
$L_1$&0.04&0.01&0.01&1.00\\
$L_2$&0.05&0.01&0.01&1.00\\
$L_3$&0.06&0.01&0.01&1.00\\
$L_4$&0.06&0.01&0.01&1.00\\
$L_\infty$&0.09&0.03&0.03&1.00\\
\noalign{\smallskip}\hline
\end{tabular}
\end{table}

Figure \ref{fig:2} compares the true distribution function $F_n$,
the Edgeworth approximation $\widehat F_n$, and the rearranged
Edgeworth approximation $\widehat F_n^*$.  The standard normal first
order approximation is also included as a benchmark of comparison.
We see that the rearranged Edgeworth approximation not only solves
the monotonicity problem, but also consistently does a better job at
approximating the true distribution than the Edgeworth
approximation. Table \ref{table1} further supports this point by
presenting the numerical results for the $L_p$ approximation errors,
calculated according to the formulas given in the previous section.
We see that the rearrangement reduces the approximation error quite
substantially in most cases.


\begin{figure*}[h]
\includegraphics[width=\textwidth,height=1.25\textwidth]{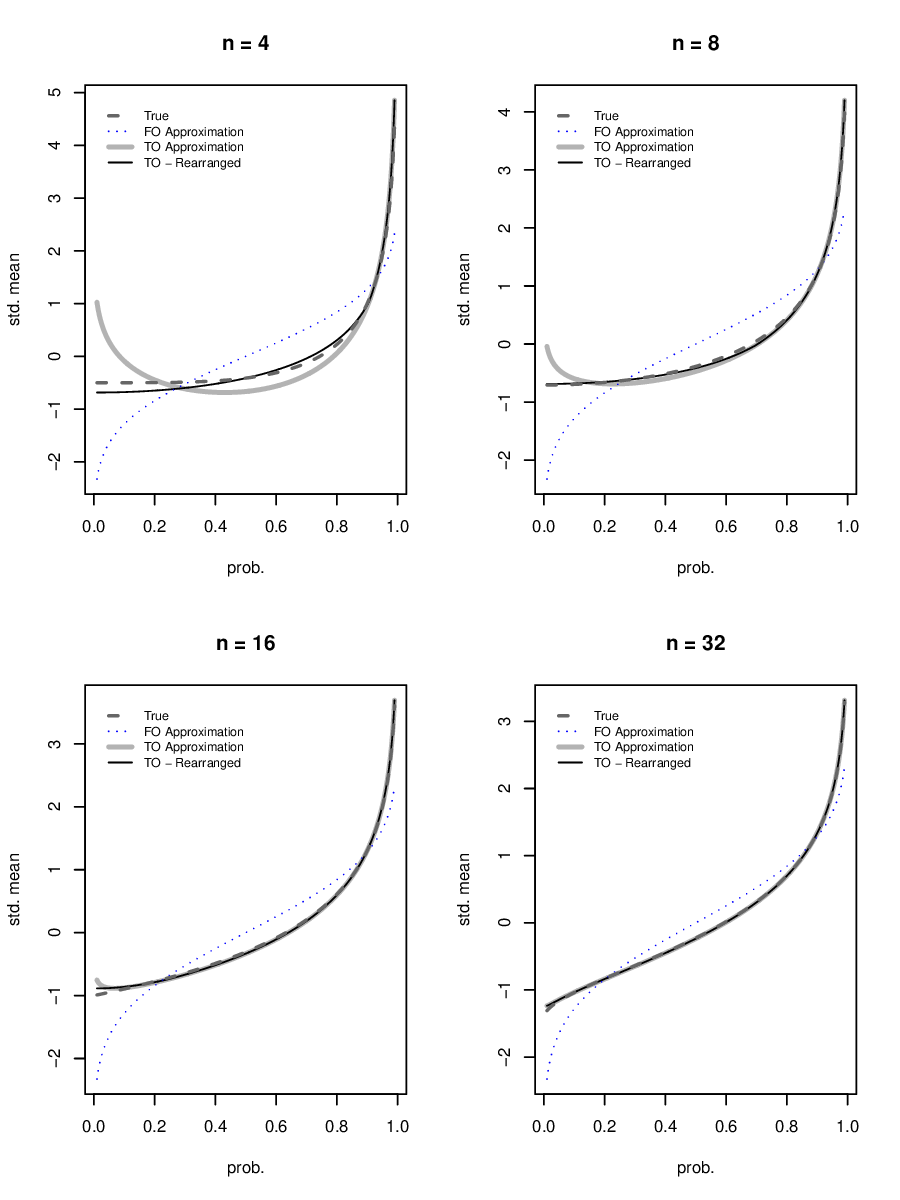}
\caption{Quantile Functions, First Order Approximations, Third Order
Approximations, and Rearrangements for the standardized sample mean
from a Gamma(1/16, 16) population.}
\label{fig:3}       
\end{figure*}


\begin{table}[h]
\caption{Estimation errors for approximations to the quantile
function of the standardized sample mean from a Gamma(1/16, 16)
population.\label{table2}}
\begin{tabular}{lllll} \hline\noalign{\smallskip}
\multicolumn{1}{l}{
}&\multicolumn{1}{c}{First Order }&\multicolumn{1}{c}{Third Order }&\multicolumn{1}{c}{Rearranged - TO}&\multicolumn{1}{c}{Ratio (RTO/TO)}\\
\noalign{\smallskip}\hline\noalign{\smallskip}
\multicolumn{1}{l}{ }&\multicolumn{4}{c}{$n = 4$}  \\
$L_1$&0.50&0.24&0.09&0.39\\
$L_2$&0.59&0.32&0.11&0.35\\
$L_3$&0.69&0.42&0.13&0.31\\
$L_4$&0.78&0.52&0.15&0.29\\
$L_\infty$&2.04&1.53&0.49&0.32\\
\multicolumn{1}{l}{ }&\multicolumn{4}{c}{$n = 8$}  \\
$L_1$&0.39&0.08&0.03&0.37\\
$L_2$&0.47&0.11&0.04&0.35\\
$L_3$&0.56&0.16&0.05&0.32\\
$L_4$&0.65&0.21&0.06&0.30\\
$L_\infty$&1.66&0.67&0.22&0.33\\
\multicolumn{1}{l}{ }&\multicolumn{4}{c}{$n = 16$}  \\
$L_1$&0.28&0.02&0.02&0.97\\
$L_2$&0.35&0.04&0.03&0.84\\
$L_3$&0.43&0.05&0.04&0.71\\
$L_4$&0.51&0.07&0.04&0.63\\
$L_\infty$&1.34&0.24&0.10&0.44\\
\multicolumn{1}{l}{ }&\multicolumn{4}{c}{$n = 32$}  \\
$L_1$&0.20&0.01&0.01&1.00\\
$L_2$&0.26&0.01&0.01&1.00\\
$L_3$&0.31&0.02&0.02&1.00\\
$L_4$&0.37&0.02&0.02&1.00\\
$L_\infty$&1.02&0.07&0.07&1.00\\
\noalign{\smallskip}\hline
\end{tabular}
\end{table}

Figure \ref{fig:3} compares the true quantile function $Q_n$, normal
first order approximation, Cornish-Fisher approximation $\widehat
Q_n$, and rearranged Cornish-Fisher approximation $\widehat Q_n^*$.
Here too we see that the rearrangement not only solves the
non-monotonicity problem, but also brings the approximation closer
to the truth. Table \ref{table2} further supports this point
numerically, showing that the rearrangement reduces the $L_p$
approximation error quite substantially in most cases.

\section{Conclusion} \label{sec:4}

In this paper, we have applied the rearrangement procedure to
monotonize Edgeworth and Cornish-Fisher expansions and any other
related expansions of distribution and quantile functions. The
benefits of doing so are twofold. First, we have obtained
approximations to the distribution and quantile curves of the
statistics of interest which satisfy the logical monotonicity
restriction, unlike those directly given by the truncation of the
series expansions. Second, we have shown that doing so results in
better approximation properties.

\begin{acknowledgements}
The results of this paper were first presented at the Statistics
Seminar in Cornell University, October 2006. We would like to thank
an anonymous referee, Andrew Chesher, James Durbin, Ivar Ekeland,
Xuming He, Joel Horowitz, Roger Koenker, Enno Mammen, Charles
Manski, Ilya Molchanov, Francesca Molinari, and Peter Phillips for
helpful discussions and comments. We would also like to thank
seminar participants at CEMMAP Microeconometrics: Measurement
Matters Conference,  Cornell University, Cowless Foundation 75th
Anniversary Conference, Transportation Conference at the University
of Columbia, University of Chicago, and University of Illinois at
Urbana-Champaign for helpful comments.
\end{acknowledgements}

\bibliography{c:/aaa/biblio/my}
\bibliographystyle{econometrica}

\end{document}